\documentclass[aps,ppr,floatfix,nofootinbib,amsmath,amssymb,amsfonts,notitlepage, twocolumn,superscriptaddress]{revtex4-2}

\usepackage{xcolor}
\usepackage{lmodern}
\usepackage{amssymb,amsmath}
\usepackage{graphicx}
\usepackage{subfigure}
\usepackage[english]{babel}
\usepackage{color}
\usepackage{booktabs}
\usepackage{blindtext}
\usepackage{comment}
\usepackage[symbol]{footmisc}

\usepackage{tikz,pgf}
\usepackage[normalem]{ulem}
\usepackage{cancel}
\usepackage[colorlinks]{hyperref}
\hypersetup{
	colorlinks=true,
	linkcolor=blue,
	citecolor=blue,
	filecolor=cyan,
	urlcolor=cyan,
}
\usepackage{notes2bib}
\usepackage{soul}
\bibnotesetup{
	note-name = ,
	use-sort-key = false
}

\begin{document}
	
	\title{Surface-Induced Spectral Broadening during Efficient Two-Photon Excitation of Cold $^{133}\text{Cs}$ Atoms near an Optical Nanofiber}
	
	\author{Yuan Jiang}
	\affiliation{State Key Laboratory of Quantum Optics Technologies and Devices, Institute of Laser Spectroscopy, Shanxi University, Taiyuan 030006, China}
	\affiliation{Collaborative Innovation Center of Extreme Optics, Shanxi University, Taiyuan 030006, People's Republic of China}
	
	\author{Xiaodong Qi}
	\affiliation{ICIQ Research, Hangzhou, China}
	
	\author{Nicolás Vera Paz}
	\affiliation{Departamento de Física, Facultad de Ciencias Físicas y Matemáticas, Universidad de Concepción, Concepción, Chile}
	
	\author{Yilin Chen}
	\affiliation{State Key Laboratory of Quantum Optics Technologies and Devices, Institute of Laser Spectroscopy, Shanxi University, Taiyuan 030006, China}
	\affiliation{Collaborative Innovation Center of Extreme Optics, Shanxi University, Taiyuan 030006, People's Republic of China}
	
	\author{Tanbin Wu}
	\affiliation{State Key Laboratory of Quantum Optics Technologies and Devices, Institute of Laser Spectroscopy, Shanxi University, Taiyuan 030006, China}
	\affiliation{Collaborative Innovation Center of Extreme Optics, Shanxi University, Taiyuan 030006, People's Republic of China}

	\author{Dianqiang Su}
	\affiliation{State Key Laboratory of Quantum Optics Technologies and Devices, Institute of Laser Spectroscopy, Shanxi University, Taiyuan 030006, China}
	\affiliation{Collaborative Innovation Center of Extreme Optics, Shanxi University, Taiyuan 030006, People's Republic of China}

	
	\author{Pablo Solano}
	\affiliation{Departamento de Física, Facultad de Ciencias Físicas y Matemáticas, Universidad de Concepción, Concepción, Chile}

	\author{Yanting Zhao}
	\email{zhaoyt@sxu.edu.cn}
	\affiliation{State Key Laboratory of Quantum Optics Technologies and Devices, Institute of Laser Spectroscopy, Shanxi University, Taiyuan 030006, China}
	\affiliation{Collaborative Innovation Center of Extreme Optics, Shanxi University, Taiyuan 030006, People's Republic of China}
	
	\date{\today}
	
	\begin{abstract}
		
		We report the experimental observation of the $^{\text{133}}$Cs 6S$_{\text{1/2}}$ to 6D$_{\text{5/2}}$ single-frequency two-photon transition at low excitation power using an optical nanofiber embedded in a cold atom cloud. We investigate the 917 nm fluorescence spectra modified by the surface-induced interaction of the nanofiber, which exhibits a pronounced asymmetric broadening with a distinct red-shifted tail. We present a systematic theoretical description of the two-photon transition process of atoms near a nanofiber surface, considering atomic angular momentum coupling, guided-mode decay rate, and surface-induced van der Waals interactions. Owing to the tight spatial confinement provided by the guided evanescent field, efficient nonlinear excitation is realized at excitation powers down to tens of microwatts, corresponding to a reduction of approximately four orders of magnitude relative to typical free-space implementations. Our work reveals the interplay among waveguide-modified radiative dynamics, surface-induced interactions, and the spatial distribution of atoms around the nanofiber, and provides insights into further studies on nonlinear optical transitions and surface-mediated atomic dynamics in waveguide quantum electrodynamics systems.

	\end{abstract}
	\maketitle
	
	\section{Introduction} 
	\raggedbottom
	Photonics platforms offer significant light–matter interactions, making them ideal for exploring nonlinear and quantum optical phenomena.\cite{luo2024strong,dutt2024nonlinear}. Optical nanofibers (ONFs), which feature strongly confined evanescent fields and extended interaction lengths, provide an efficient platform for coupling nearby atoms to guided optical fields without the need for an optical cavity \cite{tong2003subwavelength,le2005spontaneous,le2006scattering,lee2015inhomogeneous}. Over the past decade, nanofiber-based atom–light interfaces have been widely studied and employed to investigate various phenomena, including electromagnetically induced transparency (EIT) \cite{jones2015ladder,kumar2015multi,su2019observation}, electric quadrupole transitions \cite{ray2020observation}, light storage \cite{gouraud2015demonstration,sayrin2015storage}, and collective radiative effects \cite{liedl2024observation,su2023dynamical}. The strong optical confinement provided by ONFs enables higher-order processes, such as two-photon excitation, and facilitates efficient nonlinear light-matter interactions at low optical powers \cite{Gokhroo_2022,boyd2008nonlinear,vetsch2010optical,le2007spontaneous}. For example, the single-frequency two-photon transition in alkali atoms gives rise to nonlinear, intensity-dependent optical processes, such as four-wave mixing, which leads to coherent radiation through cascaded transitions from highly excited states to intermediate states and subsequently to ground states \cite{brownell1995yoked,kumar2015ats,rajasree2020spin,Hassanin2023}.

	The ONF can be used to collect light generated by such nonlinear atomic processes by coupling the emitted fields into its guided modes. However, the presence of a waveguide substantially modifies the local electromagnetic environment experienced by the atoms. The guided modes alter the local density of optical states, giving rise to Purcell-modified spontaneous emission, whereas surface-induced van der Waals (vdW) interactions shift the atomic energy levels and modify the optical spectra \cite{kien2007optical,patterson2018spectral,PhysRevA.99.013822}. Because the two-photon transition proceeds through the hyperfine manifold of the intermediate P$_{\text{3/2}}$ state, the excitation probability depends on the corresponding angular-momentum coupling coefficients.
	
	Here, we experimentally investigate 6S$_{\text{1/2}}$ to 6D$_{\text{5/2}}$ single-frequency two-photon excitation spectroscopy for cold $^{\text{133}}$Cs atoms near an ONF. The strong confinement of the evanescent field around the ONF results in a high local optical intensity, enabling efficient two-photon excitation with guided powers as low as $40~\mu\mathrm{W}$, which we study through the detection of the 917 nm fluorescence emitted from the 6D$_{\text{5/2}}$ $\rightarrow $ 6P$_{\text{3/2}}$ spontaneous decay. We developed a theoretical model incorporating angular momentum coupling, waveguide-modified spontaneous emission, and surface-induced van der Waals interactions to understand the observed spectral footprint of the surface-altered two-photon excitation. Our work reveals the interplay among waveguide-modified radiative dynamics, surface-induced interactions, and the spatial distribution of atoms around the ONF in the formation of these spectra. Understanding these mechanisms is essential for studying higher-order atomic transitions and nonlinear optics in waveguide quantum electrodynamics (QED) systems.

	\section{Experimental Setup and Results}
	The experiment consists of an ONF superimposed on a $^{\text{133}}$Cs magneto-optical trap (MOT) in a vacuum chamber. The nanofiber is 500 nm in diameter over a length of 5 mm fabricated from a standard single-mode fiber (Fibercore SM800) by the flame brushing technique \cite{hoffman2014ultrahigh,ward2014contributed}. The final diameter of the ONF is approximately 500 nm. The MOT, generated in the vacuum chamber from the residual atomic vapor pressure from a Cs dispenser, provides a source of cold atoms that couple to the evanescent field of the ONF guided mode.
	
	\begin{figure}[t]
		\centering
		\includegraphics[width=0.45\textwidth]{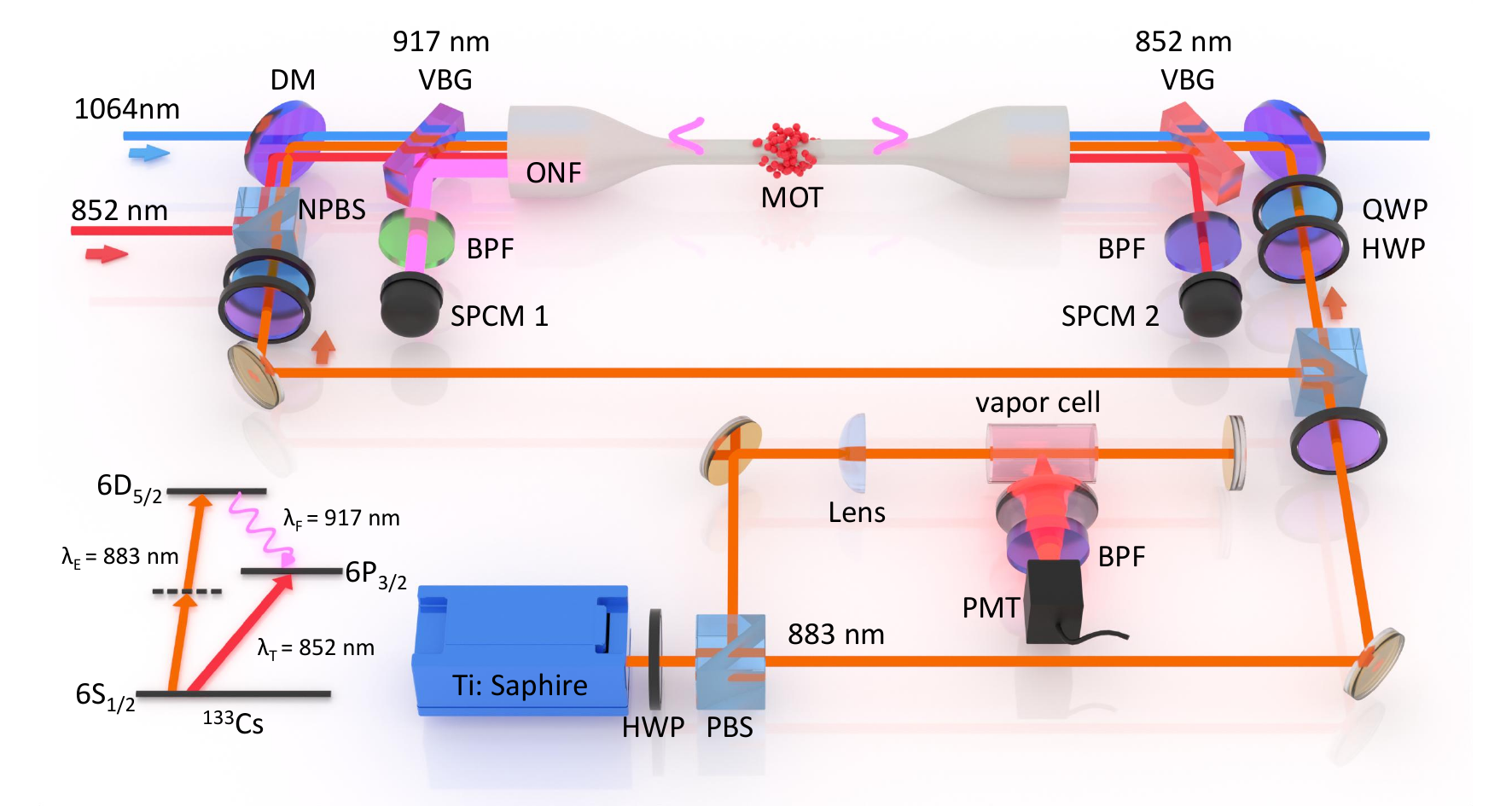}
		\caption{Experimental setup schematic  and energy-level diagram for the  $^{133}$Cs 6S$_{\text{1/2}}$ - 6D$_{\text{5/2}}$ two-photon transition. The Ti: Sapphire laser drives the 883 nm two-photon transition in Cs atoms surrounding the ONF. A 917 nm fluorescence spectrum is observed from the 6D$_{5/2}$ excited state decaying to the 6P$_{\text{3/2}}$ intermediate state. HWP: half-wave plate, QWP: quarter-wave plate, PBS: polarizing beam splitter, NPBS: 50/50 non-polarizing beam splitter, DM: dichroic mirror, BPF: bandpass filter, PMT: photomultiplier tube, SPCM: single-photon counting module. In the hot vapor cell, a frequency reference is obtained from the 852 nm fluorescence spectrum (6P$_{\text{3/2}}$ - 6S$_{\text{1/2}}$)of two-photon excitation.}
		\label{fig1}
	\end{figure}
	
	Figure \ref{fig1} shows a schematic of the experimental setup and the energy-level diagram. We first perform two-photon transition excitation in a hot vapor cell heated to 90~$^\circ \text{C}$ using counter-propagating 883 nm lasers, collecting the fluorescence signal of the $6\text{P}_{\text{3/2}} \rightarrow 6\text{S}_{\text{1/2}}$ decay with a photomultiplier tube (PMT) as a frequency reference. In the vacuum chamber, the ONF overlaps with a $^{\text{133}}$Cs MOT, which is loaded through the closed $F_4\rightarrow F_5$ D2 line transition, using a magnetic field gradient of 15 G/cm. The cold atomic cloud reaches a temperature around $\sim$130~$\mu$K. An 852 nm beam guided by the ONF, resonant with the $6\text{S}_{\text{1/2}}$ $F=4 \rightarrow 6\text{P}_{\text{3/2}}$ $F'=5$ transition in Cs, is used to verify the presence of atoms in the vicinity of the ONF by measuring the optical depth (OD) with SPCM 2.

	
	We drive the two-photon 6S$_{\text{1/2}} \rightarrow 6\text{D}_{\text{5/2}}$ transition in a cold atomic ensemble of $^{133}$Cs near the evanescent field of the ONF, using 883 nm resonant beam coupled into the ONF. For this, we employ a Ti:Sapphire laser to produce two counter-propagating excitation beams. Atoms excited to the 6D$_{\text{5/2}}$ state decay spontaneously back to the ground state via the 6P$_{\text{3/2}}$ intermediate state while emitting photons of 917 nm and 852 nm. By scanning the frequency of the 883 nm laser, the 917 nm fluorescence emission spectrum is observed (SPCM 1), which can be easily separated from the MOT light and 852 nm atomic fluorescence. Each experimental acquisition needs 60 s, and the 917 nm fluorescence spectrum is obtained by scanning the 883 nm frequency within a 50 ms window of a 100 $\mu$s bin width. We place a bandpass filter in front of the SPCM to reduce the background noise caused by Raman scattering from the ONF \cite{mitsch2014interaction}. A weak ONF-guided probe light (852 nm) is used to measure the absorption of the MOT. In addition, we inject a 1064 nm laser to boil off the excess atoms already deposited on the fiber.

\begin{figure}[t]
		\centering
		\includegraphics[width=0.35\textwidth]{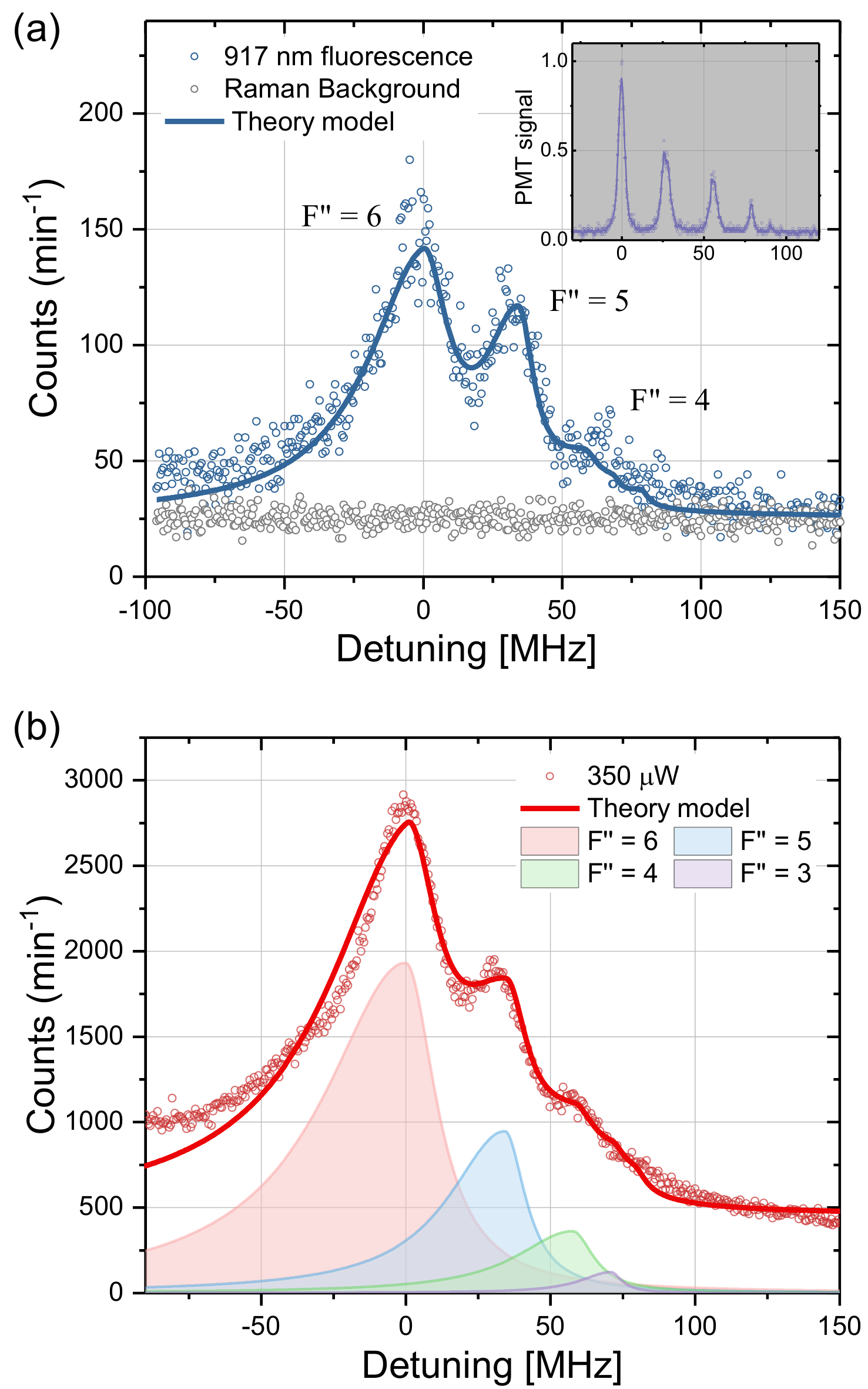}
		\caption{ Experimental demonstration of two-photon transition spectroscopy for the 6S$_{\text{1/2}}$ ground state to the 6D$_{\text{5/2}}$ excited state in cold atoms through an ONF. (a) 917 nm fluorescence signal (blue circles) detected by the SPCM with an excitation power of 40 $\mu$W. Gray circles represent the Raman scattering background. Inset: 852 nm fluorescence (6P$_{\text{3/2}}$–6S$_{\text{1/2}}$) from two-photon excitation in a hot vapor cell as a reference. (b) The fluorescence signal (red circles) of 917 nm at 350 $\mu$W with inhomogeneous broadening and a tail on the red‑detuned side. The red curve represents the simulated spectra. The colored regions represent the individual spectral components of the hyperfine energy levels in the model.}
		\label{fig2}
	\end{figure}

	
	We first characterize the two-photon excitation spectrum at low excitation power, where saturation and power broadening are weak. Figure \ref{fig2}(a) shows the 917 nm fluorescence of atomic decay from the 6D$_{\text{5/2}}$ to the 6P$_{\text{3/2}}$ state at an excitation power of P = 40 $\mu$W. The hyperfine levels $F''$ = 6, 5, and 4 of the 6D$_{\text{5/2}}$ state can be clearly observed. Simultaneously, as shown in the inset of Fig.~\ref{fig2}(a), we record a frequency reference signal of the two-photon transition at 852 nm (6P$_{\text{3/2}}$ - 6S$_{\text{1/2}}$) from the vapor cell using a PMT.
	
	As the excitation power increases, in addition to the usual power broadening, the spectrum exhibits increasingly pronounced asymmetry and inhomogeneous broadening. Figure \ref{fig2}(b) shows the fluorescence spectra recorded at higher excitation powers. At an excitation power of 350 $\mu$W, the fluorescence profile becomes noticeably asymmetric, with the red-detuned side rising to form an extended tail \cite{nayak2007optical}. These features indicate that the ONF modifies the local radiative environment experienced by the atoms surrounding the fiber, and that the measured spectrum cannot be described by a single homogeneous Lorentzian response.
	
	We introduce a unified spectral model that accounts for the hyperfine-dependent two-photon transition strength, the spatially varying evanescent-field coupling, the waveguide-modified radiative response, the cylindrical geometry of the atom–fiber shell, and the surface-induced frequency shift. The calculated fluorescence spectrum is written as
	
	\begin{align}
		S(\nu,P) &= \sum_{F''} A_{F''}\int_0^\infty \sigma_{\mathrm{eff}}(r)\,(a+r)\, \rho_{ee}(r,P)\nonumber\\
		&\!\!\!\!\cdot \! L\left(\nu \!\! -\!\! \nu_{F''} \!\! -\!\! \Delta\nu_{\mathrm{vdW}}(r); \Gamma_{\mathrm{hom}}(r,P)\right)dr \!+\! B(P),
		\label{eq:spec}
	\end{align}
	with $A_{F''} \propto S_{\mathrm{exc}}(F'')\cdot S_{\mathrm{emit}}(F'')$ the combined hyperfine amplitude for the detected spectrum of level $F''$, $S_{\mathrm{exc}}(F'')$ the excitation strength, $S_{\mathrm{emit}}(F'')$ the emission strength, $\sigma_{\mathrm{eff}}(r)$ the effective detection weight for atoms at distance $r$ from the fiber surface, $\rho_{ee}(r, P)$ the density matrix element proportional to the population of atoms at the excited state pumped by a laser of power $P$, $\Delta\nu_{\mathrm{vdW}}(r)$ the differential van der Waals frequency shifts, $L(\nu, P)$ the standard Lorentzian function of atomic spectrum, $\Gamma_{\mathrm{hom}}$ the linewidth of atoms in a homogeneous environment, and $B(P)$ the background noise contribution for the spectrum measurement, which can be easily excluded via experimental design. Throughout this paper we use two radial coordinates and keep them strictly distinct. The coordinate $r$ denotes the distance measured \emph{from the fiber surface} outwards ($r\ge 0$); the coordinate $\varrho$ denotes the distance measured \emph{from the fiber axis} ($\varrho\ge a$, with $a$ the fiber radius), so that $\varrho=a+r$. The factor $(a+r)$ in Eq.~\eqref{eq:spec} is the cylindrical Jacobian of the volume integral. Writing the volume element as $dV=\varrho\,d\varrho\,d\phi\,dz$ and changing variable to $r=\varrho-a$ gives $dV=(a+r)\,dr\,d\phi\,dz$. Because the atomic cloud is azimuthally uniform and the guided mode is treated in its azimuthal average, $\int d\phi=2\pi$; and since every quantity in the integrand depends on the radial coordinate alone, $\int dz=\mathcal{L}$ contributes a constant. Both are absorbed into the overall normalisation, leaving the single radial integral above. The $(a+r)$ factor is retained rather than approximated, because the integrand is strongly concentrated within $r\lesssim100$ nm where $(a+r)$ varies by a factor $\gtrsim 1.7$ and is therefore \emph{not} approximately constant. This expression provides the total model used to fit the measured spectra. The individual physical ingredients entering the model are described in Sec. III.
	
	\section{Theoretical model}
	\subsection{Atomic transition strengths}
	
	To understand the relative amplitudes of the observed hyperfine peaks, we first consider the transition strengths associated with the two-photon excitation process. The excitation proceeds through the intermediate $6\text{P}_{\text{3/2}}$ hyperfine manifold, giving rise to multiple quantum pathways that connect the initial and final hyperfine states. In particular, we consider the two-photon excitation pathways $6\text{S}_{\text{1/2}}(F=4) \rightarrow6\text{P}_{\text{3/2}}(F')\rightarrow 6\text{D}_{\text{5/2}}(F'')$, where the relative strength of each pathway is determined by the angular-momentum coupling coefficients, which can be expressed in terms of Wigner $6j$ symbols \cite{qi2018dispersive,PhysRevA.93.023817}. The detailed derivation of the excitation and emission strengths, including the coherent and incoherent limits of the intermediate $6P_{3/2}$ hyperfine pathways, is given in Appendix~\ref{TPES}.
	
	The actual excitation is driven by the quasi-linearly polarized HE$_{11}$ counter-propagating mode, which is not polarization-averaged \cite{snyder1983optical}. A fully $m$-resolved Clebsch-Gordan (CG) computation shows that the coherent and incoherent sums are exactly equal when summed over all photon polarizations (the polarization-averaged convention of Eq.~\eqref{eqA19}, but for linear polarization, the \textit{coherent} sum enhances $F''=5$ and $F''=3$ by approximately a factor of 2 while leaving $F''=4$ unchanged. The measured hyperfine amplitude ratio $A_5/A_6=0.72$ (low power) lies between the incoherent limit ($0.50$) and the linearly coherent limit ($0.99$), indicating partial coherence of the intermediate paths. We describe hyperfine amplitude by a single coherence parameter $\eta$ that interpolates between the two limits,
	\begin{equation}
		\begin{split}
			& A_{F''}(\eta) = A_{F''}^{\mathrm{incoh}}\left[1+\eta\,(R_{F''}-1)\right], \\
			& R_{F''} = \{5{:}1.98,\ 4{:}1.00,\ 3{:}1.97,\ 2{:}1.00\},
		\end{split}
		\label{eq-eta}
	\end{equation}
	where $\eta=0$ and $\eta=1$ correspond to the fully incoherent (polarization-averaged) and fully linearly coherent limits, respectively, and $A_{F''}^{\mathrm{incoh}}$ are the polarization-averaged ratios $A_5/A_6=0.501$, $A_4/A_6=0.216$, $A_3/A_6=0.073$.

	The coherent form exceeds the incoherent one by the interference cross terms, which for the quasi-linearly polarized HE$_{11}$ mode are constructive for $F'' = 5,3$ (Appendix~\ref{TPES}). Both strengths enter the excitation dynamics through the saturation parameter, 
	(see Eq.~\eqref{eq-app-slink}), so the partially coherent $A_{F''}(\eta)$ of Eq.~\eqref{eq-eta} sets the saturated-absorption lineshape $s/(1+s)$ of the three-level dynamics in the Appendices. This is the theoretical basis for the partial saturation analysis presented below. 
	
	
	We implement the spectral decomposition by fitting each spectrum as a superposition of asymmetric Lorentzian components (one per $F''$), whose relative amplitudes are given by Eq.~\eqref{eq-eta}. The per-$F''$ blue/red-detuned half-widths, the peak frequencies, the background and the coherence parameter $\eta$ are free parameters, subject to physically motivated constraints (monotonic amplitudes and widths, $A_{F''=4}\ge A_{F''=3}$, $\eta\in[0,1]$). The resulting best fits are summarized in Table~\ref{tab-fit} and shown in Fig.~\ref{fig2}, respectively.
	\begin{table}
		\caption{Best-fit intermediate-path coherence $\eta$ and hyperfine
			amplitude ratios at three excitation powers, from the
			partial-coherence model in Eq.~\eqref{eq-eta}. $R^2$ is the coefficient
			of the fit. The coherence decreases monotonically with
			power, indicating a coherent-to-sequential crossover of the two-photon
			excitation.}
		\label{tab-fit}
		\begin{ruledtabular}
			\begin{tabular}{lccccc}
				Power & $R^2$ & $\eta$ & $A_5/A_6$ & $A_4/A_6$ & $A_3/A_6$\\
				\hline
				40 $\mu$W & 0.907 & 0.54 & 0.765 & 0.216 & 0.110\\
				80 $\mu$W & 0.976 & 0.32 & 0.658 & 0.216 & 0.095\\
				350 $\mu$W & 0.984 & 0.00 & 0.501 & 0.216 & 0.072\\
			\end{tabular}
		\end{ruledtabular}
	\end{table}
	
	The fitted coherence parameter decreases monotonically with power ($0.54\to 0.32\to 0.00$): at low power, the two-photon process is partially coherent, and the interference between the $F'$ intermediate paths enhances $F''=5$. At high power, it becomes fully sequential/incoherent, reproducing the polarization-averaged ratios of Eq.~\eqref{eqA19}. This coherent-to-sequential crossover reflects the saturation of the intermediate state (Appendix~\ref{three-level-ladder-system-and-optical-bloch-equations}): as the single-photon Rabi frequency $\Omega\propto\sqrt{P}$ approaches the $6\mathrm{P}_{3/2}$ linewidth $\Gamma_i=2\pi\times5.2$ MHz, the intermediate level acquires a real population, its spontaneous decay destroys the phase relation between the $F'$ paths, and the two-photon event degenerates into two uncorrelated single-photon steps. The crossover power $\Omega(P_{\mathrm{cross}})\sim\Gamma_i$ corresponds to $P_{\mathrm{cross}}\sim 40$–$80\ \mu$W, consistent with the observed $\eta$ drop between 40 and 80 $\mu$W. At 350 $\mu$W, the near-surface atoms saturate and their red-shifted vdW components become visible as a broad red wing, which we include as an additional asymmetric component at $-54$ MHz with $\gamma=25$ MHz. The $F''=1$ state has zero two-photon excitation strength because $|\Delta F|>2$ is forbidden in the two-step process. In addition, the excitation strengths of the $F''=2$ and $F''=3$ transitions were extremely weak. Owing to broadening effects, they overlap with $F''=4$ and are buried in the background.
	
	
	\subsection{Surface induced phenomena}
	
	Atoms located near an ONF interact not only with radiation modes but also with the guided mode supported by a dielectric waveguide. The ONF modifies the local density of optical states, thereby changing the spontaneous emission rate of an excited atom via the Purcell effect. Consequently, the spontaneous decay rate becomes strongly dependent on the atom–surface distance and the dipole orientation. The total spontaneous emission rate can be expressed in terms of the electromagnetic Green tensor \cite{PhysRevA.70.053823,PhysRevA.93.023817} as
	
	\begin{equation}
		\frac{\gamma(\mathbf{r})}{\Gamma_0} =
		\frac{3}{2}\left(\frac{c}{\omega_0}\right)^3
		\operatorname{Im}\left[\hat{\mathbf{d}}^*\cdot
		\mathbf{G}(\mathbf{r},\mathbf{r},\omega_0)\cdot\hat{\mathbf{d}}\right],
	\end{equation}
	where $\mathbf{G}(\mathbf{r},\mathbf{r})$ is Green's tensor describing the electromagnetic response of the ONF structure, and $\Gamma_0$ is the free-space spontaneous decay rate. The Green tensor naturally includes both the radiation and guided modes supported by the ONF. For the guided mode contribution, using the Purcell factor:
	\begin{equation}
		\frac{\gamma_{\mathrm{guided}}(\mathbf{r})}{\Gamma_0} =
		\frac{\sigma_0\cdot n_g\cdot\sum_{m=1}^4|\hat{\mathbf{d}}\cdot\mathbf{E}_m(\mathbf{r})|^2}{4\cdot P_{NG}},
	\end{equation}
	where $\mathbf{E}_m(\mathbf{r})$ is the guided-mode electric field at the atomic position, $\sigma_0 = 3\lambda_0^2/(2\pi)$ is the resonant scattering cross-section in free space, $n_g = \beta/k_0$ is the group index of refraction, the sum runs over four modes (forward/backward $\times$ H/V) of the HE$_{11}$ modes of the ONF, and $P_{NG}$ is the power normalization integral. Guided modes govern the decay rate for atoms positioned close to the nanofiber surface, whereas the radiative contribution recovers the natural decay rate once the atoms are sufficiently far away. 
	Furthermore, surface-mediated nonradiative energy transfer--described by Förster and Dexter mechanisms \cite{PhysRevB.91.155313}---introduces an additional nonradiative decay channel with rate $\gamma_{\mathrm{quench}}$, given by
	\begin{equation}
		\gamma_{\mathrm{quench}}(r) = \Gamma_{q0} \cdot \exp(-r / D_{\mathrm{cut}}).
	\end{equation}
	The best fit for the cutoff distance $D_{\mathrm{cut}}$ and the peak quenching rate $\Gamma_{q0}$ are $8\sim 12$ nm and $15 \sim 20 \times \Gamma_0$, respectively. The total position-dependent transition rate of atoms near an ONF is the sum of these three components: guided, radiative, and non-radiative. 
	
	We consider atoms are allocated in a detailed balanced state with kinetic motion and rebound with the nanofiber while attracted under van der Waals potential $U_{\mathrm{vdW}}(r)$ and laser-induced potential $U_{\mathrm{dip}}(r)$, and hence the atom density follows the Boltzmann distribution:
	\begin{equation}
		p(r) \propto \exp\left[-\frac{U_{\mathrm{tot}}(r)}{k_B T}\right],
	\end{equation}
	where $T \approx 130\ \mu$K as measured, and total trapping potential is
	\begin{equation}
		U_{\mathrm{tot}}(r) = U_{\mathrm{dip}}(r) + U_{\mathrm{vdW}}(r).
	\end{equation}
	
\begin{figure}
		\centering
		\includegraphics[width=0.35\textwidth]{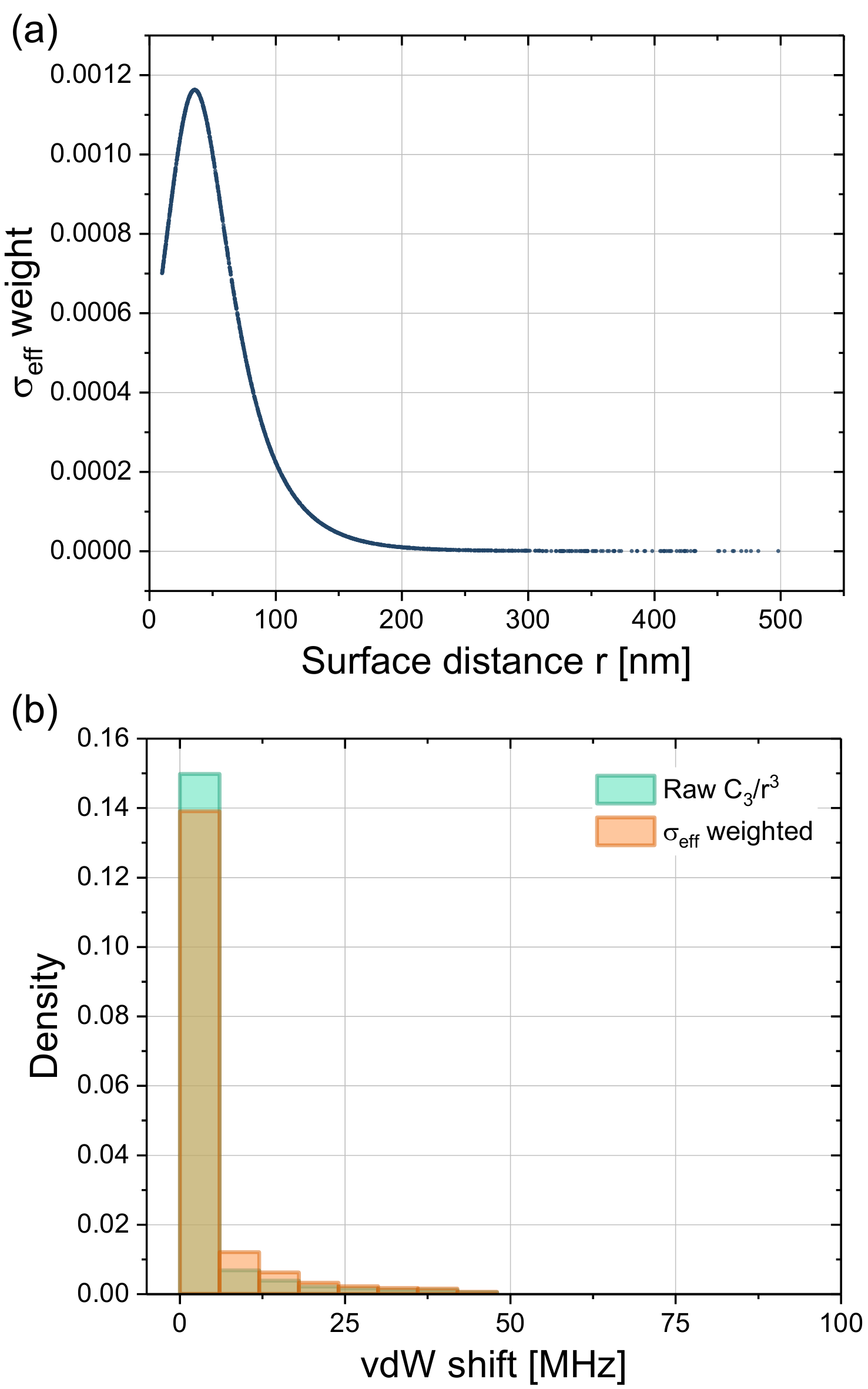}
		\caption{Surface-induced inhomogeneous broadening mechanism near the nanofiber surface. (a) Calculated effective weight $\sigma_{\text{eff}}(r)$ as a function of atom-surface distance. (b) Corresponding probability distribution of the vdW-induced frequency shifts. The $\sigma_{\text{eff}}$-weighted distribution reveals the asymmetric red-detuned broadening responsible for the observed spectral line shape.
		}
		\label{model}
	\end{figure}

	In the far-field regime, dipole trapping dominates but it decays exponentially. Near the surface ($r < 50$ nm), the vdW potential attracts atoms inward, but quenching ($\gamma_{\mathrm{quench}}$) eliminates the detectable signal from $r < 10$ nm. The measured fluorescence was collected exclusively through the nanofiber-guided mode. The effective detection efficiency for an atom at a surface distance $r$ can be formulated as:
	\begin{equation}
		\sigma_{\mathrm{eff}}(r) = p(r) \cdot f_{\mathrm{exc}}(r) \cdot
		\eta_{\mathrm{coll}}(r) \cdot \mathrm{QY}(r).
	\end{equation}
	$f_{\mathrm{exc}}(r)=I_{\mathrm{exc}}^2(r) $ is the two-photon excitation weight, which is determined by the local intensity of the excitation laser. The guided-mode
	collection efficiency $\eta_{\mathrm{coll}}(r)$ and the quantum yield $\mathrm{QY}(r)$ are respectively:
	\begin{align}
		&  \eta_{\mathrm{coll}}(r) = \frac{\gamma_{\mathrm{guided}}(r)} {\gamma_{\mathrm{guided}}(r)+\gamma_{\mathrm{rad}}(r)}, \\
		& \mathrm{QY}(r)=\frac{\gamma_{\mathrm{guided}}(r)+\gamma_{\mathrm{rad}}(r)}{\gamma_{\mathrm{guided}}(r)+\gamma_{\mathrm{rad}}(r)+\gamma_{\mathrm{quench}}(r)},
	\end{align}
	where $\gamma_{\mathrm{rad}}(r)$ is the radiative decay rate into unguided optical modes.
	In the spectral model, $p(r)$ is combined with the excitation and detection efficiencies to obtain the total effective weight (see Fig.~\ref{model}(a)):
	\begin{equation}
		\begin{split}
			& \sigma_{\mathrm{eff}}(r) = \\
			& p(r) \cdot I_{\mathrm{exc}}^2(r)  \cdot
			\frac{\gamma_{\mathrm{guided}}(r)}{\gamma_{\mathrm{guided}}(r) + \gamma_{\mathrm{rad}}(r) + \gamma_{\mathrm{quench}}(r)}.
		\end{split}	        
	\end{equation}
	The observed spectrum of the atomic ensemble should be the integration of the position-dependent $\sigma_{\mathrm{eff}}$ for atoms outside the nanofiber.


	The dominant origin of the asymmetric spectrum is the atom–surface van der Waals interaction. Atoms trapped within the evanescent-field region experience an attractive potential arising from the interaction between the fluctuating atomic dipole and dielectric ONF \cite{Frawley_2012}. In the non-retarded regime, the interaction potential can be expressed as
	\begin{equation}
		U_{\mathrm{vdW}}(r)=-\frac{C_3}{r^3},
		\label{eq4}
	\end{equation}
	where $C_3$ is the van der Waals coefficient. Since the ground and excited states experience different surface potentials, the optical transition frequency of each atom becomes distance dependent
	\begin{equation}
		\Delta\nu_{\mathrm{vdW}}(r) = \frac{C_3^{\mathrm{diff}}}{h\cdot r^3}.
		\label{eq5}
	\end{equation}
	The differential vdW coefficient computed from Lifshitz theory is $C_3^{\mathrm{diff}} = C_3(6\mathrm{D}_{5/2}) - C_3(6\mathrm{S}_{1/2}) \approx 3.5\times 10^{-49} - 1.2\times 10^{-49} = 2.3\times10^{-49}$ J$\cdot$m$^3$. This coefficient is, to leading order, identical for all hyperfine $F''$ sublevels, the residual state dependence enters through the differential polarizability and is a small correction (an effective $C_3^{(5)}/C_3^{(6)}\approx 0.5$ inferred from the power dependence of $A_6/A_5$). For $p(r)\propto\exp(-r/\lambda)$ weighted by $\sigma_{\mathrm{eff}}(r)$, the probability density distribution of the inhomogeneous vdW shift can be expressed as (see Fig.~\ref{model}(b))
	\begin{equation}
		\begin{split}
			P(\Delta\nu) \propto 
			& \Delta\nu^{-4/3}\exp\left(-\frac{1}{\lambda}\left[\frac{C_3} {h\Delta\nu}\right]^{1/3}\right) \\ 
			& \cdot\sigma_{\mathrm{eff}}\left(\left[\frac{C_3}{h\Delta\nu}\right]^{1/3}\right).		
		\end{split}
		\label{eq6}
	\end{equation}
	
	As shown in Fig.~\ref{model}(a), although atoms closer to the nanofiber experience stronger vdW shifts, their contribution is suppressed by surface quenching. Consequently, the dominant fluorescence contribution originates from atoms located at intermediate distances of $r\approx 30$–$50$ nm. The corresponding vdW shift is shown in Fig.~\ref{model}(b), giving $|\Delta\nu|\sim$ 5-40 MHz, consistent with the observed per-peak linewidth and the power-independent inhomogeneous base $\Gamma_{\mathrm{inhom}}$ of Eq.~\eqref{eq10}. 

	
	Using the peak of $F''=6$ as the frequency reference, the per-peak frequency shifts for $F''=5$ and $4$ are extracted at various powers and compared with typical thermal-atom experiments. After removing the vdW contribution, small excited-level-dependent shifts (within $\pm 6$ MHz) remained at all powers. These residual shifts do not scale linearly with intensity. A direct estimate of the AC Stark shift shows that the 883 nm guided field has $I\approx 1.8\times 10^{8}$ W/m$^{2}$ ($80\ \mu$W, effective area $A_{\rm eff}\approx 0.44\ \mu$m$^{2}$), yielding $\lesssim 0.5$ MHz for the hyperfine-resolved levels of $6\mathrm{D}_{5/2}$ at $\sim 10$ THz detuning -- is an order of magnitude too small, so AC Stark cannot explain the pattern. Therefore, we attribute the residuals primarily to the fit residual of the asymmetric lineshape and secondarily to the residual magnetic field of the MOT.
	
	Under $\sigma^{+}$ excitation and optical pumping to $m_F=4$, a geometric residual field $B_{\rm res}$ produces a shift $\mu_{\rm B}g_F m_F B_{\rm res}$ and a spread across the $m_F$ sublevels. It is important to reassess the magnitude of this contribution in light of the refined linewidth model. The power-independent inhomogeneous base $\Gamma_{\rm inhom}\approx 31.5$ MHz extracted from the linewidth analysis is dominated by the surface-vdW shift spread (the $\sigma_{\rm eff}$-weighted width of $\Delta\nu_{\rm vdW}(r)=-C_3^{\rm diff}/(h r^{3})$ over $r\approx 15$-$40$ nm amounts to $\approx 20$ MHz), not by the magnetic field. The residual field $B_{\rm res}\approx 1.1$ G inferred from the small frequency residuals contributes only $\sim 1$–$3$ MHz to the $m_F$-sublevel spread. Therefore, the magnetic field is a subdominant source of broadening, consistent with the small residual frequency shifts. The vdW interaction remains the dominant origin of the asymmetric red-detuned inhomogeneous broadening responsible for the observed lineshape.
	


	
	\begin{figure}[t]
		\centering
		\includegraphics[width=0.35\textwidth]{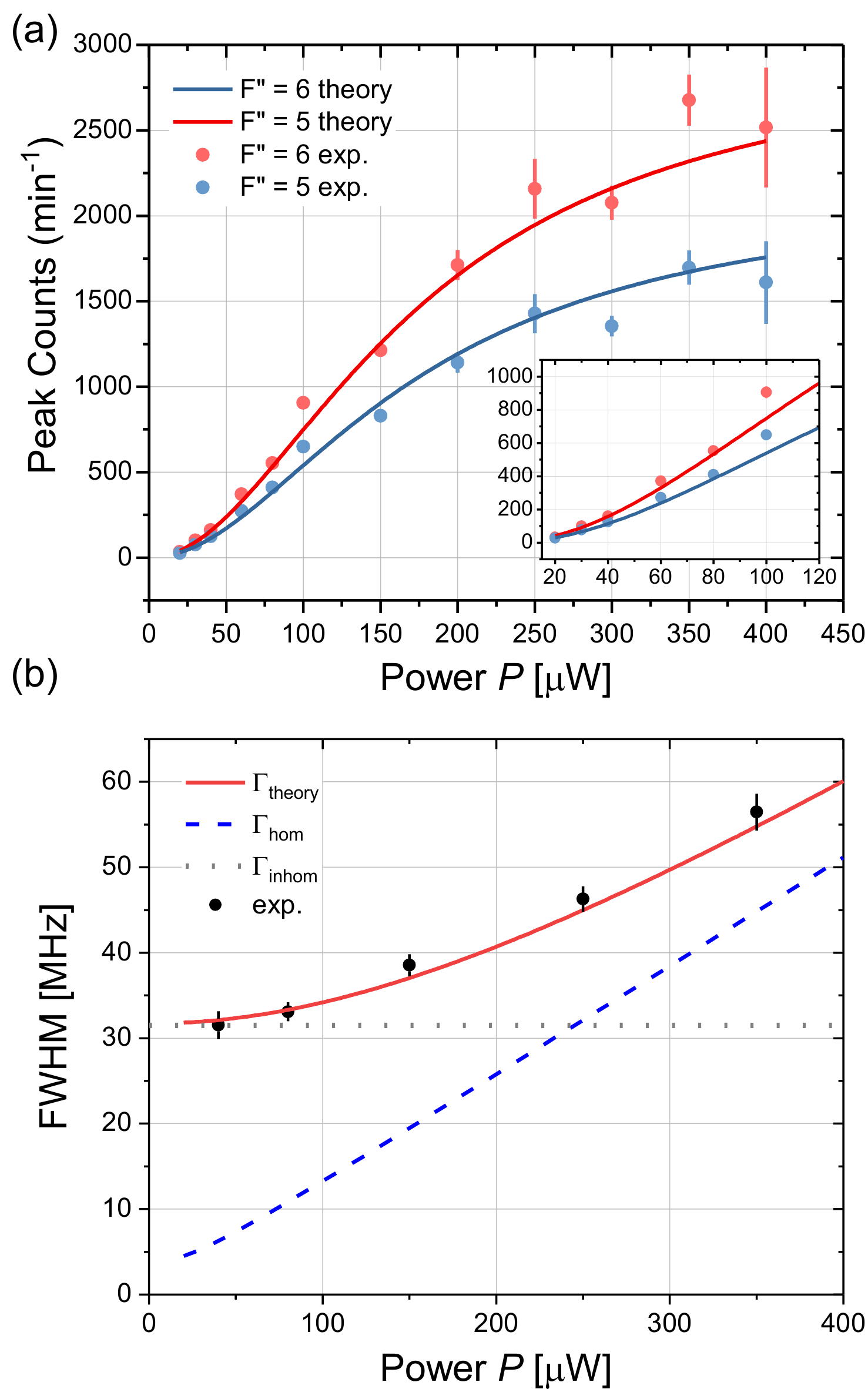}
		\caption{Spectral characteristics. (a)  Peak counts of the F$''$=6 (red) and F$''$=5 (blue) components versus power. The solid curves are the saturated-absorption (two-level) prediction $I_{F''}(P)$ with the volume-averaged $P_{\mathrm{sat}}^{\mathrm{(A)}}\approx 108\ \mu$W. The inset shows the low-power quadratic scaling $I\propto P^{2}$. The coherence parameter $\eta$ (Table~\ref{tab-fit}) decreases from 0.54 to 0 with increasing power, driving the amplified $F''$=5 component toward an incoherent ratio at high power. Error bars: standard deviation of three measurements. (b) Linewidth of the F$''$=6 fluorescence spectrum versus excitation power. Black circles represent the experiments. The blue dashed curve is the homogeneous power-broadened width $\Gamma_{\mathrm{hom}}=\Gamma_0\sqrt{1+(P/P_{\mathrm{sat}}^{\mathrm{(B)}})^{2}}$ with $P_{\mathrm{sat}}^{\mathrm{(B)}}\approx 29\ \mu$W, the gray dotted line is the power-independent inhomogeneous base $\Gamma_{\mathrm{inhom}}\approx 31.5$ MHz (surface vdW shift spread, residual MOT field, hyperfine overlap), the red solid curve is their quadrature sum, Eq.~\eqref{eq10}.}
		\label{fig3}
	\end{figure}
	
	
	\subsection{Validation of the Model with power scaling}
	
	The spatially dependent coupling strength obtained from the effective distribution also determines the different power responses of the linewidth and fluorescence intensity discussed in Fig.~\ref{fig3}. 
	We consider a single-frequency two-photon transition from the ground state $|g\rangle$ to the excited state $|e\rangle$ via a virtual level $|i\rangle$, where two counter-propagating laser beams have the same Rabi frequency, $\Omega_{ei}=\Omega_{ig}=\Omega$. The effective Rabi frequency is given by
	\begin{equation}
		\Omega_{\mathrm{eff}}=\frac{\Omega^2}{2\Delta}.
		\label{eq9}
	\end{equation}
	Because the evanescent field is spatially inhomogeneous, the saturation of the observed signal is governed by two distinct effective saturation powers, corresponding to two complementary models. In \textit{Method A} (position-averaged intensity saturation), the detected peak fluorescence is the $\sigma_{\mathrm{eff}}$-weighted ensemble average $I_{F''}(P) = A_{F''}\int \sigma_{\mathrm{eff}}(r)\,s(r,P)/[1+s(r,P)]\,dr$, which reproduces the quadratic low-power law (the inset of Fig.~\ref{fig3}(a)) and a soft (logarithmic) high-power saturation as the saturation front $r_{\mathrm{sat}}=\ln(P/P_{\mathrm{sat}})/(2q_{\mathrm{eff}})$ moves outward. This yields the volume-averaged $P^{\mathrm{(A)}}_{\mathrm{sat}}\approx 108\ \mu$W used for the peak intensities in Fig.~\ref{fig3}(a). In \textit{Method B} (near-surface saturation), the linewidth (Fig.~\ref{fig3}(b)) is dominated by the atoms at the $\sigma_{\mathrm{eff}}$ peak, where the local intensity is highest and the atoms saturate first. This yields the much smaller $P^{\mathrm{(B)}}_{\mathrm{sat}}\approx 29\ \mu$W in Eq.~\eqref{eq10}. 
	
	Because the single-photon Rabi frequency satisfies $\Omega \propto \sqrt{P}$, the effective two-photon Rabi frequency varies linearly with power, $\Omega_{\mathrm{eff}} \propto P$. The variation of the total linewidth with the excitation power is described as
	
	\begin{equation}
		\Gamma(P)=\sqrt{\Gamma'^{2}_{0}\left[1+\left(P/P^{\mathrm{(B)}}_{\mathrm{sat}}\right)^2\right]+\Gamma^{2}_{\mathrm{inhom}}},
		\label{eq10}
	\end{equation}
	where $\Gamma'_0$ is regarded as the natural linewidth of the 6D$_{\text{5/2}}$ state on the ONF surface, which is approximately equal to the spontaneous emission rate of the 6D$_{\text{5/2}}$ state in vacuum, $\Gamma'_0\approx\Gamma_0=$ 3.1 $\pm$ 0.2 MHz \cite{georgiades1994two}. In Eq.~\eqref{eq10}, $P^{\mathrm{(B)}}_{\mathrm{sat}}$ is the near-surface saturation power of Method B and $\Gamma_{\mathrm{inhom}}$ is the power-independent inhomogeneous base. We fitted the experimental data of the total linewidth using Eq.~\eqref{eq10} and obtained $\Gamma_{\mathrm{inhom}} = 31.5$ MHz and a near-surface saturation power $P^{\mathrm{(B)}}_{\mathrm{sat}} \approx 29\ \mu$W. The inhomogeneous base $\Gamma_{\mathrm{inhom}}$ combines the surface-vdW shift spread, residual MOT magnetic field (~1.1 G), and hyperfine overlap, and is power independent. The power broadening $\Gamma'_{0}\sqrt{1+(P/P^{\mathrm{(B)}}_{\mathrm{sat}})^{2}}$ is driven by the first-saturating strongly coupled atoms near the fiber surface.
	
	To make the two broadening contributions explicit, we approximate the observed linewidth by adding the homogeneous (power-broadened) and inhomogeneous (vdW + magnetic + hyperfine) contributions in quadrature, as shown in Eq.~\eqref{eq10}. This reasonable approximation avoids the use of a full Voigt convolution while capturing the experimentally observed power dependence of the linewidth. The homogeneous width follows from the effective two-level saturation parameter of Appendix~\ref{steady-state-and-saturation}, $\Gamma_{\rm hom}(r,P)=\Gamma_{\rm eff}\sqrt{1+s(r,P)}$ with $s(r,P)=(P/P_{\rm sat})^{2}e^{-4q_{\rm eff}r}$, evaluated at the $\sigma_{\rm eff}$ peak, it grows from $\sim 5$ MHz (40 $\mu$W) to $\sim 37$ MHz (350 $\mu$W) at the near-surface scale $P_{\rm sat}^{(B)}\approx 29\ \mu$W. The inhomogeneous base, $\Gamma_{\rm inhom}\approx 31.5$ MHz, is the $\sigma_{\rm eff}$-weighted width of the combined shift distribution, dominated by the surface vdW shift spread over $r\approx 15$-$40$ nm ($\approx 20$ MHz) and augmented by the residual MOT field and the overlapping $F''=4$ and $3$ hyperfine structure. Combining the two in quadrature reproduces both the zero-power intercept $\Gamma(0)=\Gamma_{\rm inhom}\approx 31.5$ MHz and the power-broadening growth, as shown in Fig.~\ref{fig3}(b). This decomposition is the spectroscopic signature of the observed linewidth saturating toward a power-broadened, absorption-like response at a high excitation power.
	
	Both Methods follow from the single local saturation parameter $s(r,P)=(P/P_{\mathrm{sat}})^{2}e^{-4q_{\mathrm{eff}}r}$, but are weighted differently: the linewidth probes the first-saturating near-surface atoms, while the intensity integrates over the whole distribution. Their ratio $P^{\mathrm{(A)}}_{\mathrm{sat}}/P^{\mathrm{(B)}}_{\mathrm{sat}} \approx 3.7$ is therefore a direct spectroscopic signature of evanescent-field inhomogeneity, where the width is set by the atoms that saturate first, while the intensity integrates over the whole distribution. This behavior is further enhanced in the ONF geometry because of the strong spatial confinement and extended interaction length of the evanescent field, which significantly increases the effective nonlinear interaction strength. Notably, owing to the strong optical confinement of the evanescent field, significant two-photon transition signals are observed at excitation powers as low as tens of microwatts, which is four orders of magnitude lower than those typically required in free-space configurations~\cite{georgiades1994two}.

	

	\section{Discussion and Conclusion}
	
	We experimentally demonstrate the single-frequency two-photon excitation of the 6S$_{\text{1/2}}$ $\rightarrow$ 6D$_{\text{5/2}}$ transition in cold $^{133}$Cs atoms near an ONF. The experimental results show that the nanofiber not only enhances the two-photon excitation efficiency but also introduces spatial selectivity to the atomic ensemble through the spatially inhomogeneous evanescent field, waveguide-modified radiation, and surface interactions. Notably, a clear two-photon signal is observed with excitation powers of only a few tens of microwatts. The fluorescence intensity and spectral linewidth exhibit distinct saturation scales, with $P_{\mathrm{sat}}^{(A)}\approx108~\mu\mathrm{W}$ and $P_{\mathrm{sat}}^{(B)}\approx29~\mu\mathrm{W}$, respectively. This difference reflects that the spectral linewidth is more sensitive to atoms located closer to the nanofiber, where the coupling is stronger and saturation occurs preferentially.
	
	The asymmetric spectral broadening further indicates that atoms close to the nanofiber surface, despite experiencing larger van der Waals shifts, contribute less to the detected fluorescence because of enhanced nonradiative quenching. Consequently, the dominant contribution to the observed signal arises from atoms at intermediate distances from the nanofiber surface. The power dependence of the hyperfine peak amplitudes also indicates a gradual transition of the two-photon excitation process from partially coherent excitation at low power toward a sequential excitation regime at higher power.
	
	Our work provides a crucial foundation for further exploration of waveguide QED of higher-order transitions. The ability to drive higher-order transitions with high efficiency makes it possible to explore multilevel atomic physics and generate non-classical light through cascaded decay processes in fiber-coupled geometries. The demonstrated high excitation efficiency opens promising avenues for exploring multilevel atomic physics, coherent frequency conversion, and the generation of nonclassical light fields via cascaded decay pathways in fiber-coupled quantum networks.

	\section*{Acknowledgments}
	We thank Luis A. Orozco for the valuable discussions and helpful comments.
	This research is funded by the National Key Research and Development Program of China (Grant No. 2022YFA1404201), and the National Natural Science Foundation of China (Grants No. 12274272, No. 12504308 ), “1331 KSC”, 111 Project (Grant No. D18001). N.V and P.S acknowledge finantial support from the National Agency for Research and Development (ANID), through Project FONDECYT Grants No. 1240204 and ANID-Subdirecci\'on de Capital Humano/Doctorado Nacional/2022-21221251.
	
	\section*{Data availability}
	The data underlying the results presented in this paper are not publicly available at this time but may be obtained from the authors upon reasonable request.
	
	\appendix 
	
	
	\section{Three-Level Ladder System and Optical Bloch Equations}\label{three-level-ladder-system-and-optical-bloch-equations}
	
	The two-photon transition proceeds through the intermediate \(6\mathrm{P}_{3/2}(F')\) manifold. We model the dynamics with the three-level ladder system \(|g\rangle\), \(|i\rangle\), \(|e\rangle\) (ground \(6\mathrm{S}_{1/2}\), intermediate \(6\mathrm{P}_{3/2}\), excited \(6\mathrm{D}_{5/2}\)). With equal Rabi frequency \(\Omega\) for both counter-propagating 883 nm legs, single-photon detuning \(\Delta = \omega_L - \omega_{ig}\) from the intermediate state, and
	two-photon detuning \(\delta = 2\omega_L - \omega_{eg}\), the Hamiltonian in the rotating frame is:
	\begin{equation}
		\begin{split}
			H=
			& -\hbar\Delta|i\rangle\langle i| - \hbar(2\Delta+\delta)|e\rangle\langle e|-  \\
			& \frac{\hbar\Omega}{2}\left(|i\rangle\langle g| + |e\rangle\langle i| + \mathrm{h.c.}\right).
		\end{split}
	\end{equation}
	Including spontaneous decay of the intermediate (\(\Gamma_i\)) and
	excited (\(\Gamma_e\)) states, the density matrix equations are:
	\begin{align}
		& \dot{\rho}_{ii} = -\Gamma_i\rho_{ii}
		+ \frac{i\Omega}{2}\left(\rho_{gi}-\rho_{ig}\right)
		+ \frac{i\Omega}{2}\left(\rho_{ei}-\rho_{ie}\right), \\
		& \dot{\rho}_{ee} = -\Gamma_e\rho_{ee}
		+ \frac{i\Omega}{2}\left(\rho_{ie}-\rho_{ei}\right), \\
		& \dot{\rho}_{ie} = -\left[i\Delta+\frac{\Gamma_i+\Gamma_e}{2}\right]\rho_{ie}
		+ \frac{i\Omega}{2}\left(\rho_{ge}+\rho_{ee}-\rho_{ii}\right), \\
		& \dot{\rho}_{ge} = -\left[i(2\Delta+\delta)+\frac{\Gamma_e}{2}\right]\rho_{ge}
		+ \frac{i\Omega}{2}\left(\rho_{ie}-\rho_{gi}\right), \\
		& \dot{\rho}_{gi} = -\left[i\Delta+\frac{\Gamma_i}{2}\right]\rho_{gi}
		+ \frac{i\Omega}{2}\left(\rho_{ii}-\rho_{gg}+\rho_{ei}\right),
	\end{align}
	with population conservation \(\rho_{gg}+\rho_{ii}+\rho_{ee}=1\) and
	conjugates \(\rho_{ij}=\rho_{ji}^*\).
	
	\section{Adiabatic Elimination of the Intermediate State}\label{adiabatic-elimination-of-the-intermediate-state}
	
	For a far-off-resonant intermediate (\(\Delta \gg \Omega, \Gamma_i\)),
	adiabatic elimination (\(\dot{\rho}_{gi}=\dot{\rho}_{ie}=0\)) yields an
	effective two-level system with $\Omega_{\mathrm{eff}}$ (effective two-photon Rabi frequency) and $\Gamma_{\mathrm{eff}}$ (effective decay, including intermediate leakage):
	\begin{align}
		& \Omega_{\mathrm{eff}} = \frac{\Omega^2}{2\Delta},  \\
		& \Gamma_{\mathrm{eff}} = \Gamma_e + \frac{\Omega^2}{4\Delta^2}\Gamma_i.
	\end{align}
	The excited-state optical Bloch equations are reduced to
	\begin{align}
		& \dot{\rho}_{ee} = -\Gamma_{\mathrm{eff}}\rho_{ee}
		- \frac{\Omega_{\mathrm{eff}}^2}{2}\left(\rho_{ge}-\rho_{eg}\right), \\
		& \dot{\rho}_{eg} = -\left(i\delta + \frac{\Gamma_{\mathrm{eff}}}{2}\right)\rho_{eg}
		- \frac{i\Omega_{\mathrm{eff}}}{2}\left(\rho_{ee}-\rho_{gg}\right).
	\end{align}
	
	\section{Steady State and Saturation}\label{steady-state-and-saturation}
	The steady-state excited population at two-photon detuning \(\delta\) is given by
	\begin{equation}
		\begin{split}
			\rho_{ee}(\delta) = \frac{s/2}{1\! +\! s \! +\! (2\delta/\Gamma_{\mathrm{eff}})^2},~s \equiv \frac{2\Omega_{\mathrm{eff}}^2}{\Gamma_{\mathrm{eff}}^2} = \frac{\Omega^4}{2\Delta^2\Gamma_{\mathrm{eff}}^2}.
		\end{split}
	\end{equation}
	
	\textbf{On resonance} (\(\delta=0\)): \(\rho_{ee}(0) = s/[2(1+s)]\), so the fluorescence rate \(R_F = \Gamma_e\rho_{ee}\) saturates as \(s/(1+s)\). Since the single-photon Rabi frequency $\Omega \propto \sqrt{P}$ and the effective two-photon Rabi frequency $\Omega_{\mathrm{eff}} \propto P$,
	\begin{equation}
		s = \left(\frac{P}{P_{\mathrm{sat}}}\right)^2.
	\end{equation}
	Here, $P_{\mathrm{sat}}$ is the local saturation power at the $\sigma_{\mathrm{eff}}$ peak, which for the power broadening of the observed linewidth coincides with the near-surface scale $P^{\mathrm{(B)}}_{\mathrm{sat}} \approx 29\ \mu$W of the main text (Sec.~III).
	
	\textbf{Power broadening.} The homogeneous linewidth grows as:
	\begin{equation}
		\Gamma_{\mathrm{hom}}(P) = \Gamma_{\mathrm{eff}}\sqrt{1+s} = \Gamma_{\mathrm{eff}}\sqrt{1 + (P/P_{\mathrm{sat}})^2}.
	\end{equation}
	In the far-off-resonant limit $\Delta \gg \Gamma_i$ the intermediate leakage term of $\Gamma_{\mathrm{eff}}$ is negligible, so $\Gamma_{\mathrm{eff}} \approx \Gamma_e = \Gamma_0$, reproducing the natural linewidth factor $\Gamma'_0$ used in Eq.~\eqref{eq10} in the main text.
	
	\section{Position-Dependent Saturation in the Evanescent Field}\label{position-dependent-saturation-in-the-evanescent-field}
	
	The guided-mode evanescent field intensity was fitted to the azimuth-averaged HE$_{11}$ mode. The fitted decay constant reproduces the real field to better than $10\%$ in every quantity reported here, including the excitation-weighted vdW width and the saturation scales, so that the exponential form is a controlled approximation rather than an assumption. In this description the modified-Bessel functions that make up the HE$_{11}$ field are evaluated at $q\varrho$, i.e. their argument uses the axis-referenced coordinate $\varrho=a+r$, \emph{not} the surface-referenced $r$. This distinction is essential rather than cosmetic: for the 500 nm fiber $q_{\mathrm{exc}}a\approx0.93$ and $q_{\mathrm{eff}}a\approx1.32$, so $qa$ is of order unity and the field cannot be written as a function of $r$ alone. Replacing $q\varrho$ by $qr$ would compress the evanescent decay length from $88$ nm to $30$ nm and shift the $\sigma_{\mathrm{eff}}$ maximum from $r\approx35$ nm down to $r\approx2$ nm, i.e. it would destroy the model. The residual azimuthal dependence of the field enters only through $\cos 2\phi$, where $\phi$ is measured from the axis of the quasi-linear polarization; it is this $\phi$ dependence that is removed by the azimuthal average. Its effective exponential decay constant is \(q_{\mathrm{eff}} = 5.3\ \mu\)m\(^{-1}\) (intensity decay length \(\approx 94\) nm, two-photon decay length \(\approx 47\) nm, and the pure-exponential \(q_{\mathrm{exc}}=3.74\ \mu\)m\(^{-1}\) form underestimates the decay because of the modified Bessel prefactors). The two-photon weight is therefore \(I^2(r) \propto e^{-4q_{\mathrm{eff}}r}\) and the local saturation parameter is:
	\begin{equation}
		s(r,P) = \left(\frac{P}{P_{\mathrm{sat}}}\right)^2 e^{-4q_{\mathrm{eff}}r},
	\end{equation}
	with local excited population $\rho_{ee}(r,P) = s(r,P)/[2(1+s(r,P))]$ and local homogeneous width
	$\Gamma_{\mathrm{hom}}(r,P) = \Gamma_{\mathrm{eff}}\sqrt{1+s(r,P)}$.
	
	\section{Two-Photon Excitation Strength: Coherent and Incoherent Forms}\label{TPES}
	
	The two-photon excitation strength $S_{\mathrm{exc}}(F'')$ depends on
	whether the intermediate $F'$ paths add coherently or incoherently. 
	To make this explicit, we first write the reduced dipole matrix elements of the hyperfine transitions via the Wigner--Eckart theorem,
	\begin{equation}
		\begin{split}
			\langle F'||d||F_g\rangle &= (-1)^{F'+J_g+I+1}\sqrt{(2F'+1)(2J_g+1)}\\
			&\quad\times\begin{Bmatrix}J_g & F_g & I\\F' & J_p & 1\end{Bmatrix}\langle J_p||d||J_g\rangle,
		\end{split}
		\label{eq-we1}
	\end{equation}
	\begin{equation}
		\begin{split}
			\langle F''||d||F'\rangle &= (-1)^{F''+J_p+I+1}\sqrt{(2F''+1)(2J_p+1)}\\
			&\quad\times\begin{Bmatrix}J_p & F' & I\\F'' & J_d & 1\end{Bmatrix}\langle J_d||d||J_p\rangle,
		\end{split}
		\label{eq-we2}
	\end{equation}
	The two-photon path through the intermediate hyperfine level $F'$
	carries the amplitude
	\begin{equation}
		A_{F'} = \frac{\langle F''||d||F'\rangle\,\langle F'||d||F_g\rangle}{\Delta_{F'}},
		\label{eq-Ap}
	\end{equation}
	where $\Delta_{F'}=\omega_{L}-\omega_{F'F_g}$ is the single-photon
	detuning from intermediate level $F'$. The $6\mathrm{P}_{3/2}$
	hyperfine splitting is $\sim 150$ MHz whereas
	$\Delta_{F'}\approx -12.5$ THz at two-photon resonance, so the relative
	hyperfine variation of the denominator is
	$\delta\Delta_{F'}/\Delta \sim 150\,\mathrm{MHz}/12.5\,\mathrm{THz}
	\sim 1.2\times 10^{-5}$. The detuning is therefore, to better than
	$10^{-4}$ accuracy, $F'$-independent, $\Delta_{F'}\approx\Delta$, and
	does not distort relative hyperfine amplitudes. This quantitatively
	justifies factoring a single $\Delta$ out of the excitation strength.
	
	\textbf{Incoherent (diagonal) limit.} When no phase relation links the
	intermediate paths the excitation strength is the sum of the squared
	path amplitudes,
	\begin{equation}
		\begin{split}
			S_{\mathrm{exc}}^{\mathrm{incoh}}(F'') &= \sum_{F'}|A_{F'}|^{2}\\
			&= \frac{1}{\Delta^{2}}\sum_{F'}
			\left|\langle F''||d||F'\rangle\,\langle F'||d||F_g\rangle\right|^{2}.
		\end{split}
		\label{eq-app-incoh}
	\end{equation}
	Substituting Eqs.~\eqref{eq-we1}-\eqref{eq-we2} and summing over the
	$m$-sublevels of $F'$ (via the Wigner-$6j$ orthogonality relations)
	reproduces the closed form 
	\begin{equation}
		\begin{split}
			S_{\mathrm{exc}}^{\mathrm{incoh}}(F'') & \propto \frac{(2J_p+1)(2J_g+1)(2F''+1)}{\Delta^{2}}\\ & \!\!\!\!\!\!\!\!\!\!\!\!\!\!\!\!\cdot\sum_{F'} (2F'+1)
			\left| \begin{Bmatrix} J_g & F_g & I \\ F' & J_p & 1 \end{Bmatrix}
			\begin{Bmatrix} J_p & F' & I \\ F'' & J_d & 1 \end{Bmatrix} \right|^2,
		\end{split}
		\label{eqA19}
	\end{equation}
	where the electronic angular momentum $J_g=1/2$, $J_p=3/2$, $J_d=5/2$, the nuclear spin $I=7/2$ for $^{133}$Cs. $F_g=4$ is the optically pumped ground hyperfine level. 
	Equation~\eqref{eqA19} is written as proportionalities because it retains only the angular-momentum (geometric) part and suppresses the common reduced dipole matrix elements; the omitted factor is $|\langle J_p||d||J_g\rangle|^{2}|\langle J_d||d||J_p\rangle|^{2}$, which does not depend on $F''$. This is immaterial for the analysis below, which uses only the normalized weights $A_{F''}/A_6$, but it makes the status of the corresponding expressions unambiguous.
	The prefactor follows directly from Eqs.~\eqref{eq-we1}--\eqref{eq-we2}: the two Wigner--Eckart square roots carry $\sqrt{(2F'+1)(2J_g+1)}$ and $\sqrt{(2F''+1)(2J_p+1)}$, so squaring the product of the two reduced matrix elements and folding the $(2F'+1)$ into the sum over $F'$ leaves the degeneracy factor $(2J_g+1)(2J_p+1)(2F''+1)$ together with the $1/\Delta^{2}$ already present in Eq.~\eqref{eq-app-incoh}. Eq.~\eqref{eqA19} is therefore the closed form of Eq.~\eqref{eq-app-incoh} with the geometric combination $|6j_{1}6j_{2}|^{2}$ displayed explicitly.
	
	The emission strength for each $F''$ via the cascaded 6$\text{D}_{\text{5/2}} \rightarrow 6\text{P}_{\text{3/2}}$ is
	\begin{equation}
		\begin{split}
			S_{\mathrm{emit}}(F'') & \propto (2J_p+1)(2F''+1)\\
			& \cdot \sum_{F'=2}^{5}(2F'+1) \left| \begin{Bmatrix} J_d & F'' & I \\ F' & J_p & 1 \end{Bmatrix} \right|^2,
		\end{split}
		\label{eqA20}
	\end{equation}
	where the sum over $F'$ is fixed by the Wigner-$6j$ orthogonality relation to an $F''$-independent value, so that $S_{\mathrm{emit}}(F'')\propto(2F''+1)$. This is the $F''$ dependence that drives the emission-side weights.
	and the combined detection weight is the product
	\begin{equation}
		A_{F''}^{\mathrm{incoh}} \propto S_{\mathrm{exc}}^{\mathrm{incoh}}(F'') \cdot S_{\mathrm{emit}}(F'').
		\label{eqA21}
	\end{equation}
	Coupling with the emission strength gives the detection weights, normalized to $F''=6$,
	\begin{equation}
		A_{F''}^{\mathrm{incoh}}\!/\! A_6 \!=\!
		\{F''\!\!{=}5{:}0.501, 4{:}0.216, 3{:}0.073, 2{:}0.015\},
		\label{eq-app-Aincoh}
	\end{equation}
	
	\textbf{Coherent (interference) limit.} When the intermediate paths
	retain a phase relation, the amplitudes add \emph{before} squaring,
	\begin{equation}
		\begin{split}
			S_{\mathrm{exc}}^{\mathrm{coh}}(F'') &= \left|\sum_{F'}A_{F'}\right|^{2}\\
			&= \frac{1}{\Delta^{2}}\left|\sum_{F'}\langle F''||d||F'\rangle\,\langle F'||d||F_g\rangle\right|^{2}.
		\end{split}
		\label{eq-app-coh}
	\end{equation}
	Expanding the square separates the two contributions,
	\begin{equation}
		S_{\mathrm{exc}}(F'') =
		\!\!\!\!\!\!\!\!\!\!\!\!\underbrace{\sum_{F'}|A_{F'}|^{2}}_{\text{diagonal (incoherent)}}
		\!\!\!+\underbrace{2\sum_{F'<F''}\mathrm{Re}\!\left(\frac{c_{F'}c_{F''}^{*}}{\Delta^{2}}\right)}_{\text{interference (coherent)}},
		\label{eq-app-expand}
	\end{equation}
	with $c_{F'}=\langle F''||r||F'\rangle\langle F'||r||F_g\rangle$. The
	diagonal term is precisely the incoherent strength of
	Eq.~\eqref{eq-app-incoh}, where the second term is the coherent gain. This is
	the explicit relation between the two limits: the coherent strength
	equals the incoherent term plus the interference cross term. Summed
	over \emph{all} photon polarizations $q_1,q_2$ the cross terms cancel
	identically by angular-momentum orthogonality, so
	$S_{\mathrm{exc}}^{\mathrm{coh}}=S_{\mathrm{exc}}^{\mathrm{incoh}}$ (the
	polarization-averaged limit of Eq.~\eqref{eqA19}). For linear
	polarization (approximately, the quasi-linearly-polarized HE$_{11}$ guided mode)
	the interference is constructive for $F''=5,3$ and negligible for
	$F''=4,2$, giving the enhancements $R_{F''}$ of Table~\ref{tab-sexc}.
	The resulting amplitude ratios and interference enhancement are
	summarized in Table~\ref{tab-sexc}.
	
	\begin{table}
		\caption{Hyperfine amplitude ratios for the two limits and the
			linearly-polarized interference enhancement
			$R_{F''}=A_{F''}^{\mathrm{coh}}/A_{F''}^{\mathrm{incoh}}$.}
		\label{tab-sexc}
		\begin{ruledtabular}
			\begin{tabular}{lcccc}
				$F''$ & 5 & 4 & 3 & 2\\
				\hline
				$R_{F''}$ & 1.98 & 1.00 & 1.97 & 1.00\\
				$A_{F''}^{\mathrm{incoh}}/A_6$ & 0.501 & 0.216 & 0.073 & 0.015\\
				$A_{F''}^{\mathrm{coh}}/A_6$ & 0.99 & 0.216 & 0.144 & 0.015\\
			\end{tabular}
		\end{ruledtabular}
	\end{table}
	
	The intermediate-path coherence is then interpolated between the two
	limits with the single parameter $\eta$,
	\begin{equation}
		A_{F''}(\eta) = A_{F''}^{\mathrm{incoh}}\left[1+\eta\,(R_{F''}-1)\right],
		\label{eq-app-eta}
	\end{equation}
	or Eq.~\eqref{eq-eta} in the main text. Here $\eta=0$ reproduces
	the fully incoherent (polarization-averaged) ratios and $\eta=1$ the
	fully linearly-coherent ones, the measured power-dependent $\eta$
	(Table~\ref{tab-fit}) quantifies the coherent-to-sequential crossover.
	
	\textbf{Relation to the saturation dynamics.} 
	The excitation strength enters the two-photon saturation parameter of
	Appendix~\ref{steady-state-and-saturation}. There the resonance population
	and the saturation parameter are expressed in terms of the effective
	two-photon Rabi frequency $\Omega_{\mathrm{eff}}=\Omega^{2}/(2\Delta)$,
	with $\rho_{ee}(0)=s/[2(1+s)]$ and $s=2\Omega_{\mathrm{eff}}^{2}/
	\Gamma_{\mathrm{eff}}^{2}$. The single-photon Rabi frequency satisfies
	$\Omega\propto\langle F'||d||F_g\rangle E/\hbar$, because the two-photon
	process sums over the intermediate manifold, $\Omega_{\mathrm{eff}}$
	is built coherently from the path amplitudes
	($\Omega_{\mathrm{eff}}\propto\sum_{F'}A_{F'}$) in the coherent limit,
	whereas in the incoherent (sequential) limit the population fed to
	$|e\rangle$ accumulates as $\sum_{F'}|A_{F'}|^{2}$. Hence
	\begin{equation}
		s(r,P)\propto\Omega_{\mathrm{eff}}^{2}\propto P^{2}\,
		S_{\mathrm{exc}}(F''),
		\label{eq-app-slink}
	\end{equation}
	Therefore, the partial saturation lineshape $s/(1+s)$ that governs the spectra of Fig.~\ref{fig3} is set by the coherent or incoherent excitation strength according to the power-dependent coherence $\eta$. This is the theoretical basis for the saturated absorption description used throughout the analysis.

	\bibliography{ref}

\end{document}